\documentstyle[12pt,epsf,fleqn]{elsart}

\def \ni{\noindent}

\def \be {\begin{equation}}
\def \ee {\end{equation}}
\def \HM {HEIDEL\-BERG-MOSCOW~}
\def \GTF {GENIUS-Test-Facility~}

\begin{document}

\ni {\bf Corresponding author}\\
Prof. Dr. H.V. Klapdor-Kleingrothaus\\
Max-Planck-Institut f\"ur Kernphysik\\
Saupfercheckweg 1\\
D-69117 HEIDELBERG\\
GERMANY\\
Phone Office: +49-(0)6221-516-262\\
Fax: +49-(0)6221-516-540\\
email: klapdor@gustav.mpi-hd.mpg.de\\

\begin{frontmatter}
\title{First Ten\,kg of Naked Germanium Detectors in Liquid Nitrogen 
	installed in the GENIUS-Test-Facility}

\author{H.V. Klapdor-Kleingrothaus}
\footnote{Spokesman of HEIDELBERG-MOSCOW and GENIUS Collaborations,\\
E-mail: klapdor@gustav.mpi-hd.mpg.de,\\ 
 Home-page: $http://www.mpi-hd.mpg.de.non\_acc/$} 
\protect\newline {O. Chkvorez, I.V. Krivosheina, H. Strecker, C. Tomei}

\address{Max-Planck-Institut f\"ur Kernphysik, PO 10 39 80, 
  D-69029 Heidelberg, Germany}

\date{18.05.2003}

\begin{abstract}

	The first four naked high purity Germanium detectors 
	were installed successfully in liquid nitrogen 
	in the GENIUS-Test-Facility (GENIUS-TF) 
	in the GRAN SASSO Underground Laboratory on May 5, 2003.
	This is the first time ever that this novel technique 
	aiming at extreme background reduction in search 
	for rare decays is going to be tested underground. 
	First operational parameters are presented.

\end{abstract}
\end{frontmatter}

\section{Introduction}

	The present status of cold dark matter search, 
	of investigation of neutrinoless double beta decay 
	and of low-energy solar neutrinos all require 
	new techniques of {\it drastic} reduction of background 
	in the experiments. For this purpose we proposed 
	the GENIUS (GErmanium in liquid NItrogen Underground Setup) 
	project in 1997 
\cite{KK-Bey97,HVKK-Prop87,Hel-KK-Genius97,KK-Hir97-Genius,KK-H-H-97,CERN87}.
	The idea is to operate 'naked' Ge detectors in liquid nitrogen, 
	and thus, by removing all materials from the immediate 
	vicinity of the Ge crystals, to reduce the background 
	considerably with respect to conventionally operated detectors. 
	The liquid nitrogen acts both as a cooling medium and 
	as a shield against external radiactivity.

	That the removal of material close to the detectors 
	is the crucial point for improvement of the background, 
	we know from our experience with the \HM double beta decay 
	experiment
\cite{KK02-Found}, 
	which is the most sensitive double 
	beta experiment for 10 years now.
	Monte Carlo simulations for the GENIUS project, 
	and investigation of the new physics potential 
	of the project have been performed in great detail, 
	and have been published elsewhere 
\cite{KK-H-H-97,GEN-prop}.
	Already in 1997 it has been shown experimentally in our Heidelberg 
	low-level facility (shielding $\sim$ 10 mwe) that 
	the techniques of operating 'naked' Ge detectors 
	in liquid nitrogen is working and we were the first 
	to show that such device can be used for spectroscopy 
\cite{KK-H-H-97}.

	A small scale version of GENIUS, the \GTF has been 
	approved by the Gran Sasso Scientific Committee in March 2001. 
	The idea of GENIUS-TF is to prove the feasibility 
	of some key constructional features of GENIUS, such as detector 
	holder systems, achievement of very low thresholds 
	of specially designed Ge detectors, long term stability 
	of the new detector concept, reduction of possible noise 
	from bubling nitrogen, etc.

	Additionally the  GENIUS-TF will improve 
	the limits on WIMP-nucleon cross
	sections with respect   
	to our results  with the HEIDELBERG-MOSCOW and HDMS experiments
\cite{prd-hdmo,HDMS03} 
	thus allowing for a test of the claimed
	evidence for WIMP dark matter from the DAMA experiment 
\cite{dama3}. 
	The relatively large mass of Ge in the full scale GENIUS-TF 
	compared to existing
	experiments would permit to search directly for a WIMP signature in
	form of the  predicted 
\cite{freese} 
	seasonal modulation of the event rate 
\cite{NIM-Mod03}. 
	Introducing the strongly 'cooled down' enriched detectors of the
	\HM $\beta\beta$-experiment into the GENIUS-TF setup,
	may allow, in a later stage, to improve the present accuracy of the
	effective Majorana neutrino mass determined recently 
\cite{KK02,KK02-PN,KK02-Found}.
	A detailed description of GENIUS-TF project is given in 
\cite{NIM02-TF}.

\begin{figure}[htp]
\epsfysize=75mm\centerline{\epsffile{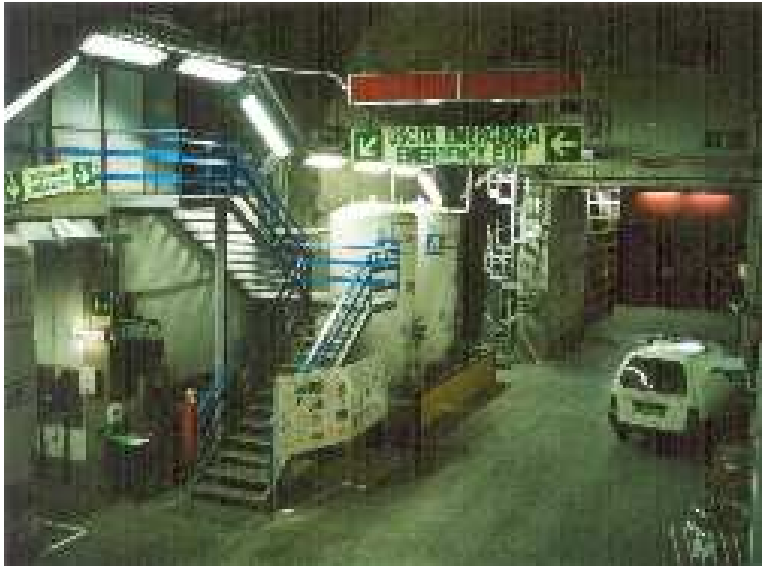}}
\caption[]{ 
	Location of GENIUS-TF is the building 
	on the right (car in front), opposite to the \HM experiment 
	building (left side).}
\label{fig:Bahnhof}
\end{figure}

	After installation of the GENIUS-TF setup between halls A 
	and B in Gran Sasso, opposite to the buildings of the \HM 
	double beta decay experiment and of the DAMA experiment (Figs. 
\ref{fig:Bahnhof}, \ref{fig:Filling-Syst-TF}), 
	the first four detectors have been installed 
	in liquid nitrogen on May, 5 2003 and have started operation.


	This is the first time ever, that this novel technique 
	for extreme background reduction in search for rare decays 
	is tested under realistic background conditions in an underground 
	laboratory.

	In section 2 we will describe the actual setup, including 
	the measures taken for producing high-purity nitrogen, 
	the measurement system of the liquid nitrogen level, 
	the new digital data acquisition system
\cite{NIM02-TF-Elektr}, 
	and will present first 
	measured spectra. In section 3 we give a short Summary.


\section{Description of Setup and of Present Performance}

\begin{figure}[htp]
\epsfysize=65mm\centerline{\epsffile{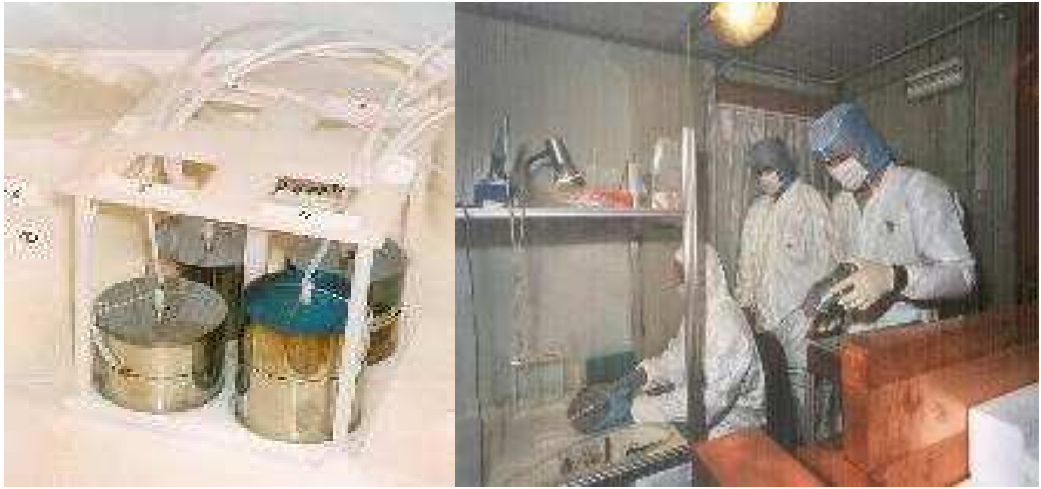}}
\caption[]{\underline{Right:} 
	Taking out the crystals from the transport dewars 
	and fixing the electrical contacts in the clean room 
	of the GENIUS-TF building - from left to right: 
	Herbert Strecker, Hans Volker Klapdor-Kleingrothaus, Oleg Chkvorez. 
	\underline{Left:} The first four contacted naked Ge detectors 
	before installation into the GENIUS-TF setup.} 
\label{fig:Foto-4det}
\end{figure}

	On May 5, 2003 the first four naked Ge detectors 
	were installed under clean-room conditions 
	into the GENIUS-TF setup. Fig. 
\ref{fig:Foto-4det}
	shows the contacted crystals after taking them out 
	of the transport dewars, in the holder made 
	from high-purity PA5 (a type a teflon), in which they then 
	are put into liquid nitrogen. 
	Each detector has a weight of 2.5\,kg. The depth of the core 
	of the detectors was reduced to guarantee a very low threshold, 
	estimated by ORTEC to be around 0.5-0.7\,keV, 
	with only marginal deterioration of the energy resolution.


\begin{figure}[htp]
\epsfysize=80mm\centerline{\epsffile{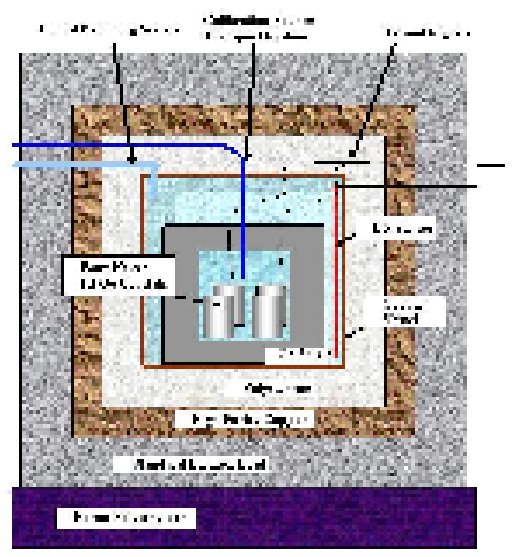}}

\vspace{.3cm}
\epsfysize=60mm\centerline{\epsffile{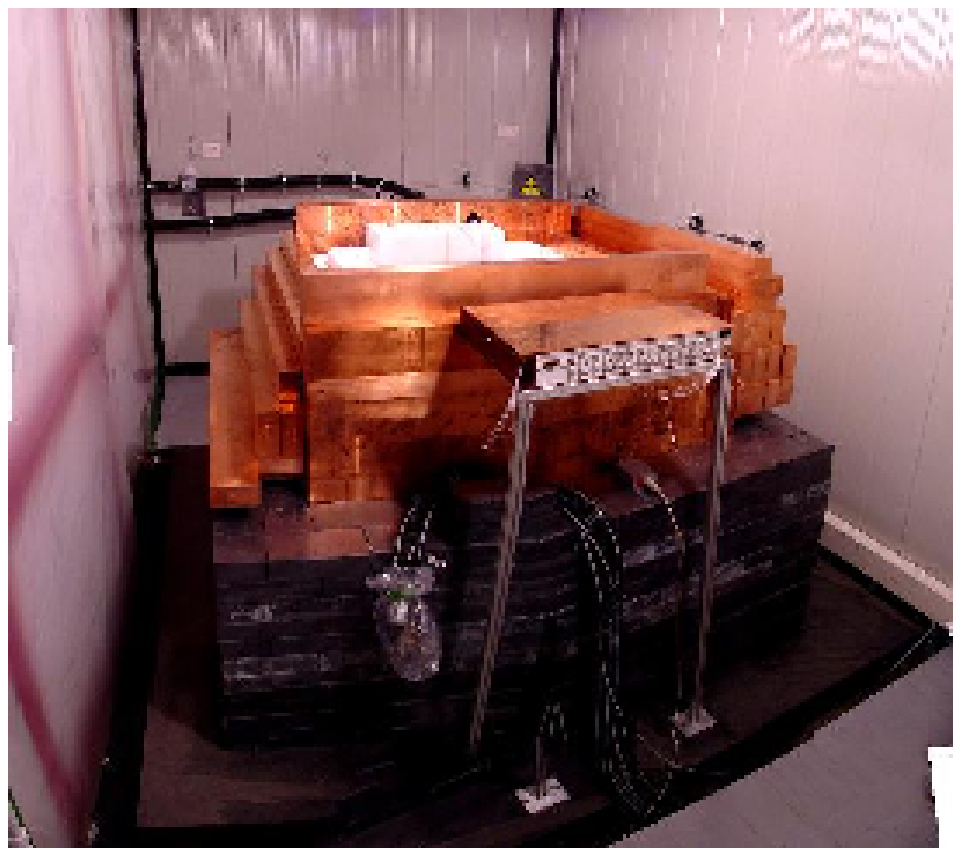}}
\caption[]{ View of GENIUS-TF in the Gran Sasso Underground Laboratory 
	in Italy.
	\underline{Upper part:} Cross section of the setup.  
	\underline{Lower part:} 
	The setup with detectors inside, but shielding only 
	partly mounted. In front the preamplifier system.} 
\label{fig:Geom-G-TF}
\end{figure}

	The liquid nitrogen (in total $\sim$ 70\,l) is kept 
	in a thin-walled (1\,mm) box of high-purity electrolytic 
	copper of size 50x50x50 cm$^3$. Inside this copper box, 
	i.e. also inside the liquid nitrogen, is installed another 
	box with walls 
	of 5 - 10\,cm monocristalline Ge bricks ($\sim$300\,kg) 
	forming the first highly efficient shield 
	of the Ge detectors (see Fig. 
\ref{fig:Geom-G-TF}).


	The copper box is thermally shielded by 20\,cm 
	of special low-level styropor, followed by a shield of 10\,cm 
	of electrolytic copper (15\,tons) and 20\,cm of low-level 
	(Boliden) lead (35\,tons). Fig. 
\ref{fig:Geom-G-TF}
	shows the geometry of the setup, and also the setup 
	in the status of not yet fully closed copper and lead shields. 
	The setup will finally by shielded against neutrons 
	with 10\,cm Boronpolyethylene plates.

\begin{figure}[htp]
\epsfysize=50mm\centerline{\epsffile{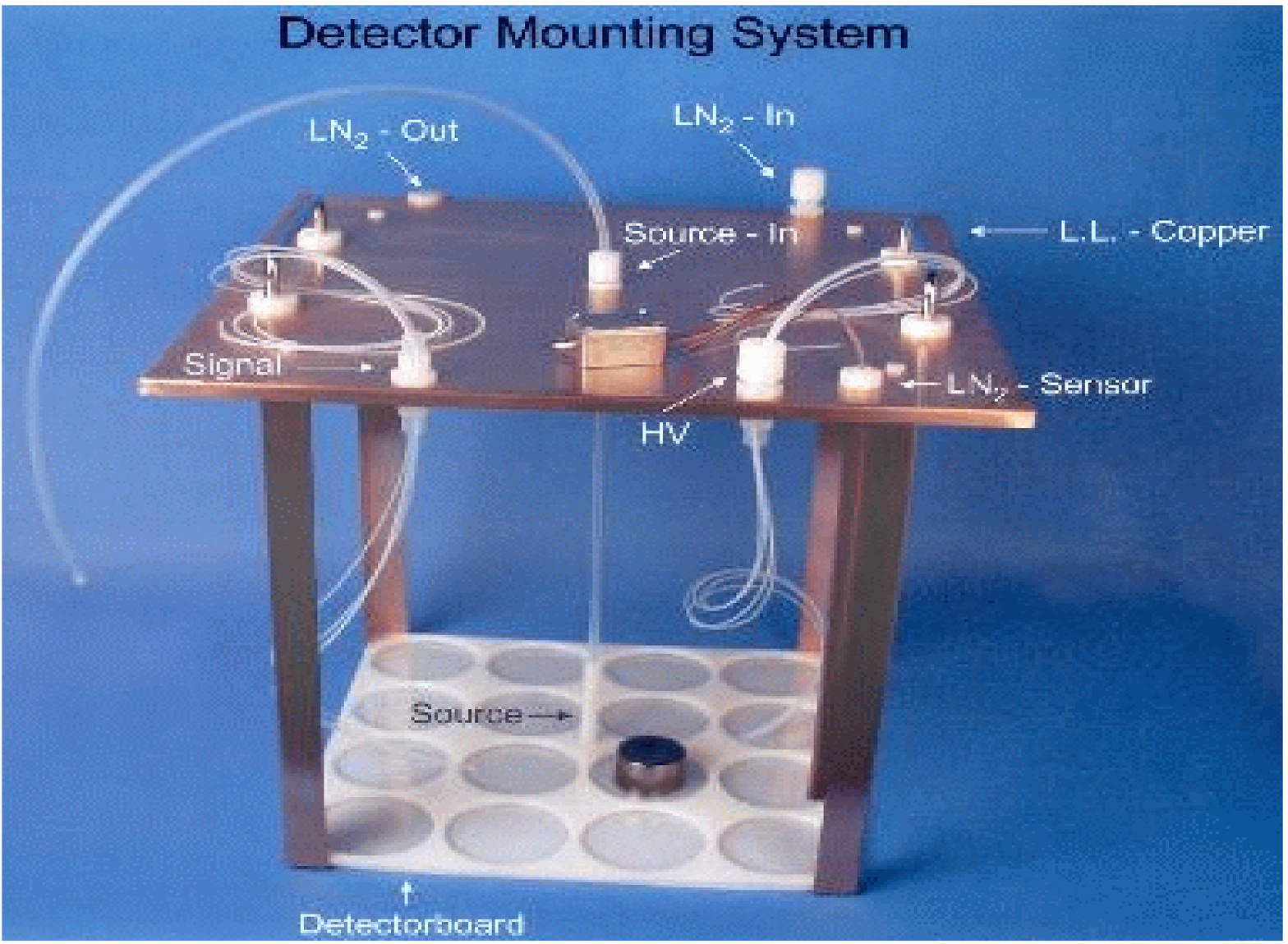}}
\caption[]{Connections of electronics, liquid nitrogen, 
	source and LN$_2$ sensor to the inner part of GENIUS-TF.} 
\label{fig:New-LiN-schem}
\end{figure}

	The high-purity liquid nitrogen used, is produced 
	by the BOREXINO nitrogen plant, which has been extended 
	for increase of the production capacity to be able 
	to provide enough nitrogen also for GENIUS-TF. 
	Liquid nitrogen of standard 
	quality (99.99$\%$ purity) is directly purified in the liquid 
	phase by an adsorber column system, consisting of 
	two independent columns (Low Temperature Adsorber - LTA) 
	filled with about 2 kg of 'activated carbon' each. 
	One of them we purchased to supplying GENIUS-TF. 
	The system is designed to continuously produce about 150 l 
	of liquid nitrogen per hour, respectively about 100 m$^3$/h 
	gaseous nitrogen for both experiments. During the regeneration 
	phase of one column the other one is in use. 
	The plant is shown in Fig. 
\ref{fig:New-LiN-schem}.
	For the experimentally measured strong reduction of Rn 
	by the cryogenic column adsorption see  
\cite{DPGAachen03}.

\begin{figure}[htp]
\epsfysize=70mm\centerline{\epsffile{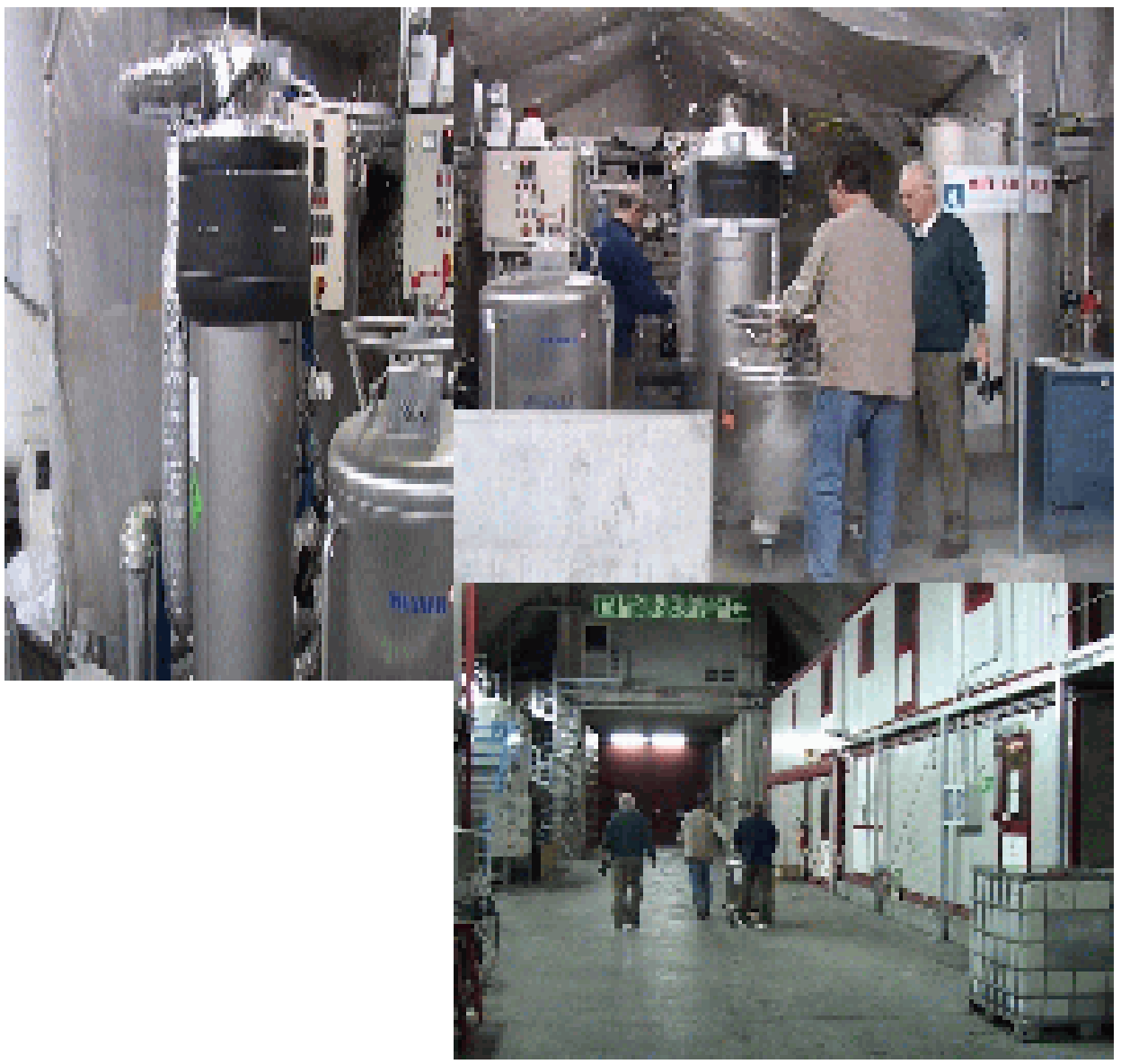}}
\caption[]{ BOREXINO-GENIUS-TF nitrogen purification system in GRAN SASSO 
	(left and upper right). 
	The left part shows the absorber column 
	(low temperature absorber - LTA) provided by the GENIUS-TF group.
	The nitrogen is transported by 200\,l vessels to the GENIUS-TF 
	building (lower right).}
\label{fig:Filling-Syst-TF}
\end{figure}  

	From the production plant the liquid nitrogen is transported by 200\,l 
	vessels to the building of the experiment. 
	Filling of the copper container with liquid nitrogen 
	is provided by connecting them to the filling system 
	consisting of isolated teflon tubes as shown in (Fig. 
\ref{fig:Filling-Syst-TF}).
	The nitrogen level in the detector chamber is measured 
	by a capacitive sensor consisting of two 40\,cm 
	long isolated selected-material copper tubes, 
	one inside the other. The change of the medium between 
	the tubes by the entering liquid nitrogen leads to a change 
	of the capacity, which is measured by subsequent electronics 
	and indicated by LED's outside of the setup. We measure 
	the nitrogen level in ten steps between 0 and 100\%. 
	GENIUS-TF has to be refilled every two days (with 
	some reserve of one more day).

	The data acquisition system we developped recently 
	for GENIUS-TF and GENIUS is decribed in detail in 
\cite{NIM02-TF-Elektr}.
	It uses multichannel digital processing technology 
	with FLASH ADC modules with high sampling rates of 100\,MHz 
	and resolution of 12\,bits. 
	It allows to capture the detailed shape of the preamplifier 
	signal with high-speed ADC, and then perform digitally 
	all essential data processing functions, including precise 
	energy measurement over a range of 1 keV - 3\,MeV, 
	rise time analysis, ballistic deficit correction 
	and pulse shape analysis. Thus we obtain both the energy 
	and the pulse shape information from one detector 
	using one channel of the Flach ADC module. 

	To allow for regular calibration of the detectors, 
	a source of $^{133}{Ba}$ fixed on a wire can be 
	introduced through a teflon tube into the center among 
	the detectors.
	The source is transported via a magnetic system.
	The activity of the source is 401\,kBq.


\begin{figure}[htp]
\epsfysize=80mm\centerline{\epsffile{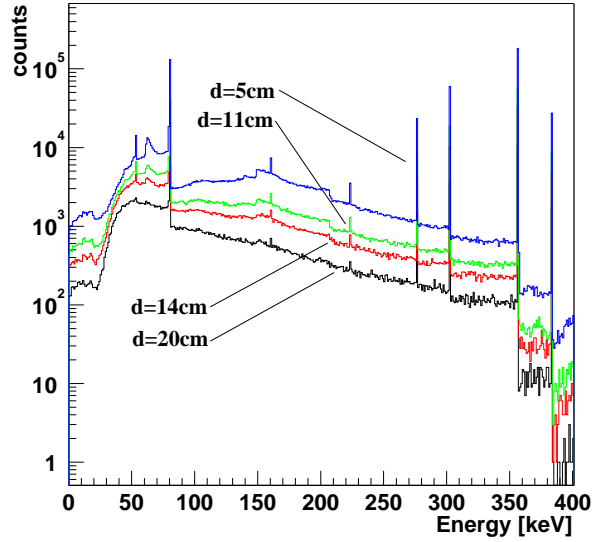}}
\caption[]{Monte Carlo simulation of GENIUS-TF calibration measurements with a
  	movable $^{133}{\rm{Ba}}$ source, for different source-detector 
	distance, with GEANT4. {\bf d} is the vertical distance 
	of the source from the plane, on which the detectors are sitting.} 
\label{fig:det1}
\end{figure}  

	Fig. 
\ref{fig:det1} 
	shows the dependence of the expected spectrum seen 
	by the four detectors as function of the position 
	in the setup ($d$ is the vertical distance of the source 
	from the plane, on which the detectors are lying. 
	For $d\sim$ 7-10\,cm the source is approximately 
	on top of the detectors).

\begin{figure}[htp]
\epsfysize=80mm\centerline{\epsffile{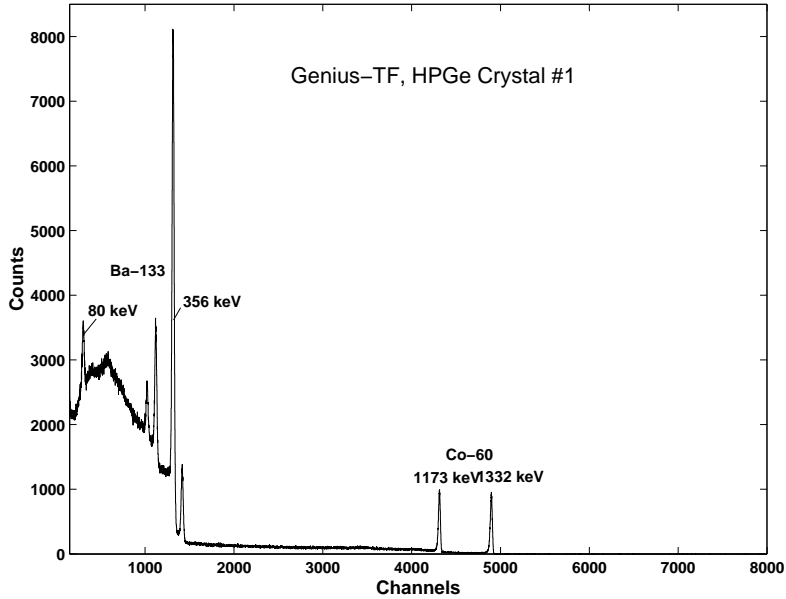}}
\caption[]{ A first spectrum measured with detector 1 with a $^{60}{Co}$ 
	source outside, and the $^{133}{Ba}$ source inside the setup.}
\label{fig:FirstSpect-G-TF}
\end{figure}

\begin{figure}[htp]
\epsfysize=80mm\centerline{\epsffile{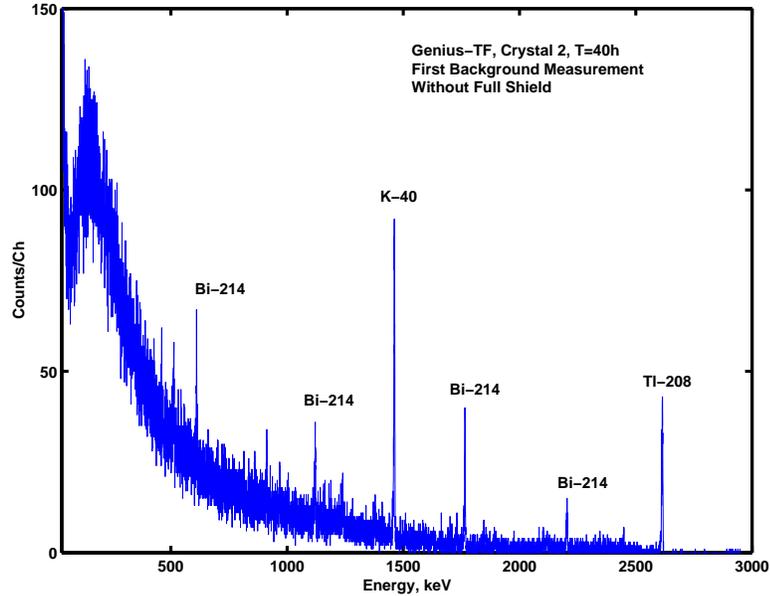}}
\caption[]{ The first background spectrum measured with 
	detector 2 over 40\,hours without shield of the setup to the top.}
	\label{fig:firstbackgr}
\end{figure}

	Figs. 
\ref{fig:FirstSpect-G-TF}, \ref{fig:firstbackgr}
	show two spectra measured a few days after installation. 
	A first spectrum measured with a $^{60}{Co}$ source 
	{\it outside} the setup, and the $^{133}{Ba}$ 
	source {\it inside}, is shown in Fig. 
\ref{fig:FirstSpect-G-TF}.
	The resolution at this moment (two days after installation)
	is 3\,keV in the 1330\,keV region.


	Fig. 
\ref{fig:firstbackgr}
	shows the background, measured with the still {\it open} 
	setup to the top. 
	When the liquid nitrogen level {\it decreases}, the background 
	slightly {\it increases}. This shows that the radioactive 
	impurities seen (from $^{40}{K}$, and the $^{232}{Th}$ 
	and $^{238}{U}$ natural decay chains) are located {\it outside} 
	the setup. No cosmologically produced impurities 
	in the detectors are seen on the present level of sensitivity.

	The effect of microphonics from bubbling in the liquid 
	nitrogen is as far as it can be seen now, negligible 
	for high energies, but has to be discriminated by pulse 
	shape analysis for low energies.
	This can be done by the new digital data acquisition system 
\cite{NIM02-TF-Elektr}.


\section{Conclusion}

	The first four naked 
	Ge detectors (10\,kg) have been installed 
	in liquid nitrogen into the \GTF in the GRAN SASSO 
	Underground Laboratory on May 5, 2003. 
	This is the first time that this novel technique  
	is applied under realistic background conditions 
	of an underground laboratory. With the successful start 
	of GENIUS-TF a historical step has been achieved 
	into a new domain of background 
	reduction in underground physics in the search for 
	rare events - and at the same time, a first experimental 
	step to GENIUS. 
	Besides testing of constructional parameters for the 
	GENIUS project one of the first goals of GENIUS-TF will 
	be to test the signal of cold dark matter reported 
	by the DAMA collaboration 
\cite{dama3},
	which could originate from modulation of the WIMP flux 
	by the moving of the Earth relative to that of the Sun.

\section{Acknowledgment:}

	The authors would like to thank their colleagues from the BOREXINO 
	collaboration for generously allowing us to use their 
	liquid nitrogen purification system, and G. Heusser 
	for his help in extending this plant for GENIUS-TF. 
	They also thank their colleagues 
	from Kurchatov Institute Moscow, for providing 
	the monocrystalline Germanium.


\end{document}